# A wide field corrector with loss-less and purely passive atmospheric dispersion correction

Peter Gillingham*, Will Saunders

Australian Astronomical Observatory, PO Box 915, North Ryde, NSW 1670, Australia

## ABSTRACT

A 2.5 degree field diameter corrector lens design for the Cassegrain focus of the VISTA 4 meter telescope is presented. It comprises four single elements of glasses with high UV transmission, all axi-symmetric for operation at the zenith. One element is displaced laterally to provide atmospheric dispersion correction. A key feature, especially beneficial for the VISTA application, is that the ADC element can be mounted so it is driven simply by gravity; thus its operation needs no motors, encoders, cabling, or software control. A simple mechanical design to achieve this and the optical performance details are described.

## 1. INTRODUCTION

For the 4MOST project, it is intended to mount a multi-fiber feed for spectroscopy at the Cassegrain focus of the VISTA 4 meter telescope at Cerro Paranal. A wide field corrector is needed to provide imaging across a field diameter of 2.5 degrees with the exit pupil concentric with the spherical focal surface. A baseline optical design was produced within the project collaboration in 2012 using a standard atmospheric dispersion corrector (ADC) comprising two doublets with inclined interfaces between the two glasses, which rotate around the optical axis.

Alternatives involving lateral motion of one doublet, rather than two rotating doublets, have been employed, e.g. in the Subaru Suprime Cam corrector and there has been a proposal[1] for a Cassegrain corrector employing a laterally displaced singlet. Recently, one of us (WS) developed an all silica design[2] for a prime focus corrector, with a very effective ADC action involving two elements moving laterally and rotating slightly around a transverse axis. This work led to the present design for VISTA, which improves on the optical resolution of the 2012 design as well as having better blue transmission, lighter glass, and a much simpler ADC action .

## 2. OPTICAL DESIGN

### 2.1 Layout

Figures 1 and 2 show, respectively, the detailed layout of the corrector design and its relation to the primary and secondary mirrors of the telescope. The second element acts as the ADC, being offset laterally by up to about 9 mm for the largest design zenith angle, 55°. The glasses are, from the front, LLF1, LLF1, N-FK5, and N-FK5. With the central thicknesses as shown (75, 55, 50, and 60 mm) the total mass of the finished elements is 310 kg and the transmission loss through internal absorption through the four elements is as plotted in figure 3 with 5% loss through most of the wavelength range, rising to 10% at the lower limit.

### 2.2 Imaging performance at zenith

Considering the asymmetry when the ADC element is offset, the array of field positions shown in figure 4 was needed to adequately characterize the performance. Figure 5 shows spot diagrams and included energy plots at the zenith.

*pg@aao.gov.au

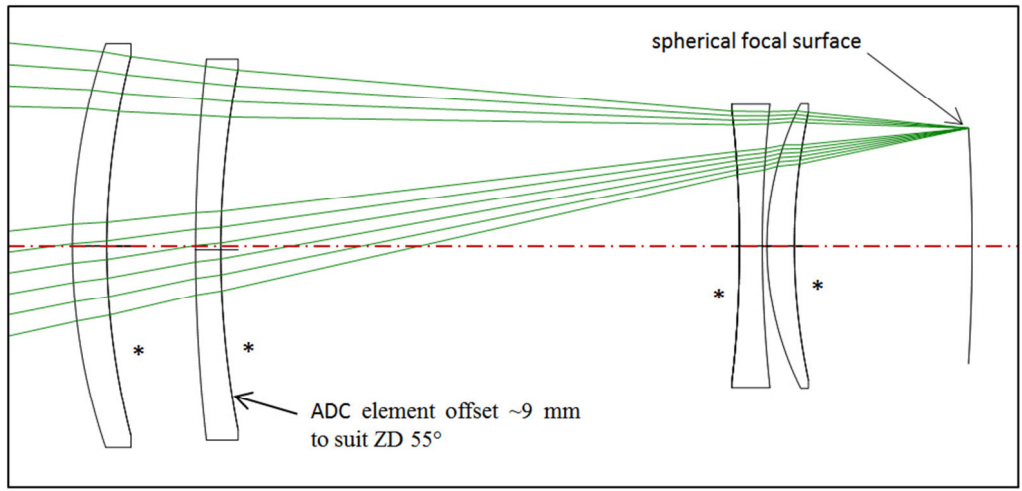

Figure 1. Layout of corrector. The second element gives the ADC action through translation downward. The surfaces marked * are polynomial aspherics.

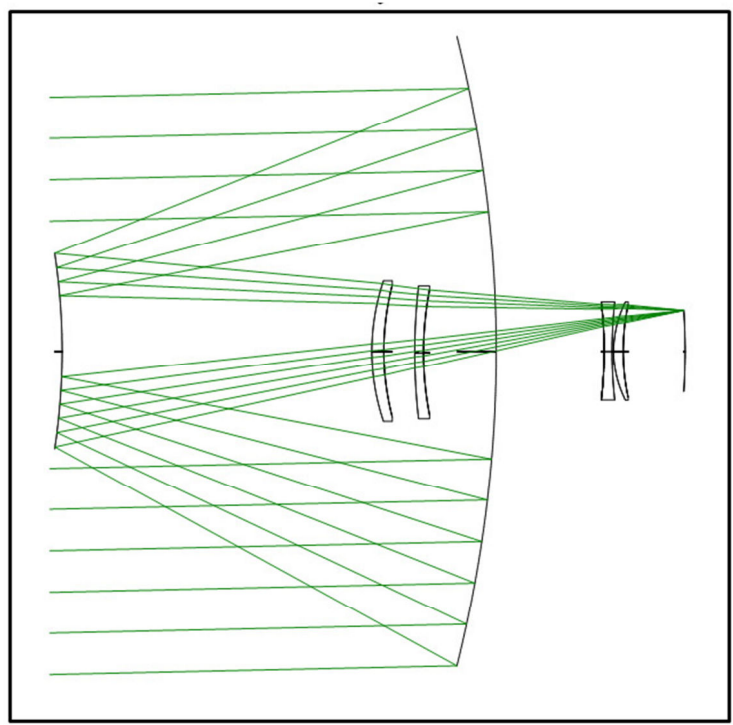

Figure 2. Relationship of corrector lens to primary and secondary telescope mirrors.

**Atmospheric dispersion correction**

The 4MOST concept selection document[3] includes the following: "The distribution of air masses expected was determined from the current VISTA IR-imaging surveys, given that 4MOST can be expected to observe similar targets.

The current VISTA survey air mass distribution peaks at air mass 1.1, with a long tail to air mass 1.7, with a median air mass of 1.25 (zenith distance of ~37.5 degree)."

Based on this, the ADC has been designed for operation at up to ZD 55° (air mass ~ 1.74), with optical tests in addition at ZD 0, 20°, and 38°.

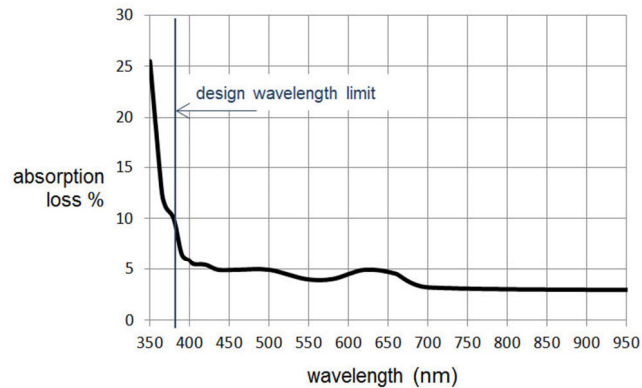

Figure 3. Absorption loss through the four elements as a function of wavelength.

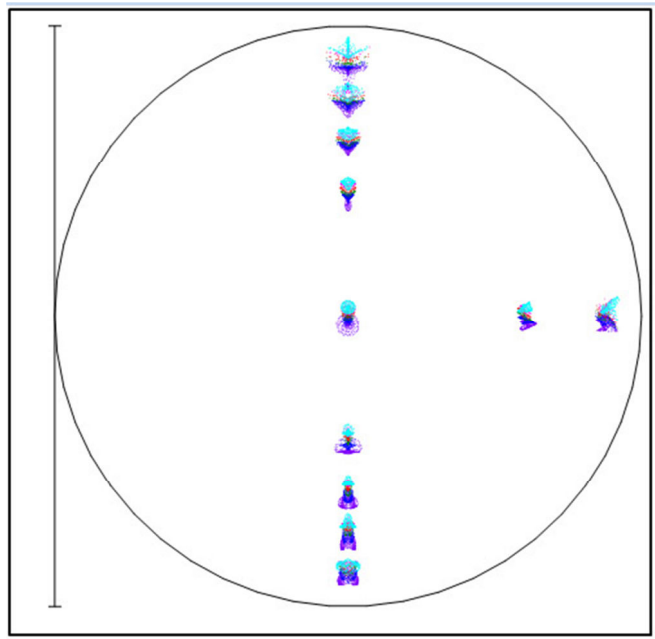

Figure 4. Full field spot diagram showing the fields for which ray-tracing was done. With the asymmetry introduced by the ADC, it was necessary to sample the field equally above and below the X axis. The aberrations (for ZD 55°) are exaggerated by a linear factor 500 and the circle diameter represents 580 mm. The four Y field radii are chosen to enclose ¼, ½, ¾, and all the field area. The two X field radii enclose ½ and all the area.

It is proposed that the offset of the ADC lens be purely passive, driven by gravity against spring restraints. This will be simplest to implement if the offsets are made proportional to sin (ZD) i.e. proportional to the lateral component of gravity, since the restraint force can then be linear with deflection. This implementation will be designated **SS** (for Single Spring). However, the atmospheric dispersion is close to proportional to tan (ZD) and the discrepancy is significant at the upper end of the ZD range 0 to 55°. The optimum ADC offset at ZD 55° is 9.6 mm while with offsets proportional to sin(ZD), an appropriate compromise offset at 55° is 6.8 mm. A slightly more complicated mechanical arrangement (**DS** for Double Spring) gives a restraining force roughly proportional to tan (ZD), resulting in a compromise offset at 55° of 8.4 mm. The optical performances with these two arrangements will be compared quantitatively.

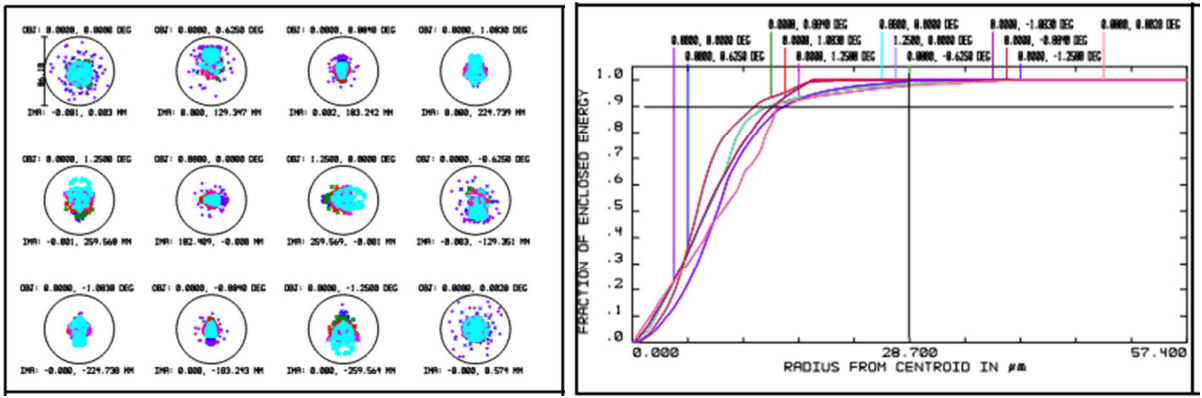

Figure 5. Spot diagrams and plots of enclosed energy for all fields at the zenith. The equally weighted wavelengths are 380, 430, 500, 600, 800, and 1000 nm. The circle diameters for the spots are equivalent to 1.5 arcsec (a likely diameter for the fiber cores). In the enclosed energy plots, the intersection of the horizontal and vertical lines is at 90% within 1 arsec diameter.

There can be some advantage optically in tilting the ADC element proportional to its offset; i.e. rotating it about an axis perpendicular to the telescope optical axis. However, this further complicates the mechanical design and it has been concluded that the performance with pure translation is adequate.

### 2.3 ADC optical performance

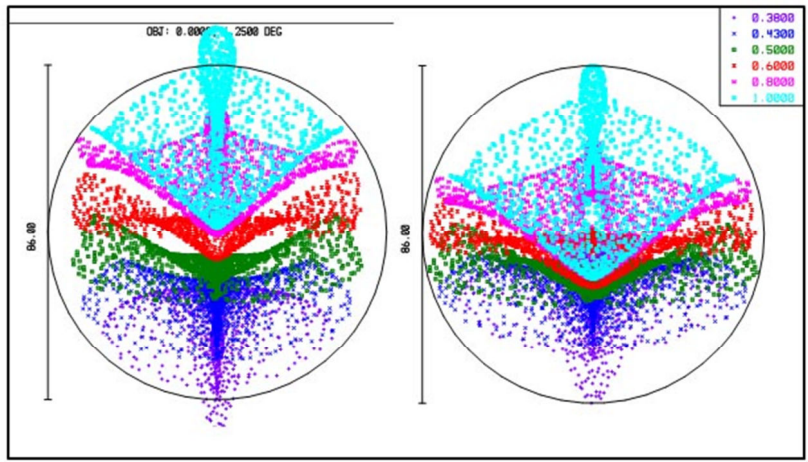

Figure 6. Spot diagrams for the worst images (at the field periphery ) at ZD 55° with: left, ADC restraint **SS**, proportional to sin(ZD) ; right: the more elaborate arrangement **DS**, approximating restraint proportional to tan(ZD). The wavelengths (μm) are indicated at the top right. The circle diameters are equivalent to 1.5 arcsec.

Figure 6 compares the spot diagrams at the worst ZD 55° field position with the simple ADC offsetting (**SS**) with that for the more elaborate version (**DS**). Clearly the former suffers from more residual atmospheric dispersion. To quantify the resulting penalty in efficiency when feeding optical fibers with core diameters equivalent to 1.5 arcsec, the Zemax simulation illustrated in figure 7 was made. The "object" was a 21 by 21 array of numbers from 0 to 9 representing a Moffat function. This was scaled to represent seeing with FWHM 0.7 arcsec. To establish the collection efficiency into a fiber, a dummy surface was placed in contact with the focal surface, having an aperture diameter equivalent to 1.5 arcsec, at a field position corresponding to the center of the relevant spot. Then the numbers of rays passing the aperture for different wavelengths were compared with the numbers with a much larger aperture. For case DS, the efficiencies were 81 and 82% at the wavelength extremes (380 and 1000 nm, respectively). For the simpler mechanical case **SS**, the corresponding efficiencies were 71 and 63%. Thus, with mechanism SS, ~12% and ~23% are lost at the blue and red limits of the wavelength range compared with mechanism DS, for ZD 55°. The relative loss would naturally be less for intermediate wavelengths and for smaller zenith angles.

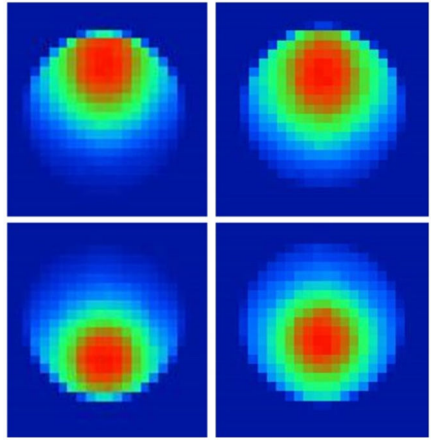

Figure 7. False color simulated images with seeing 0.7 arcsec FWHM for the worst field position with ZD 55°, top: with wavelength 1000 nm; below: with wavelength 380 nm. Left: for the simple mechanism **SS**; right: for the double spring mechanism **DS**. The images are vignetted by an aperture equivalent to 1.5 arcsec in diameter at field positions to compromise between the red and blue wavelengths.

Figure 8 shows spot diagrams for the four zenith distances tested, with the ADC mechanism **DS**. Deterioration in image quality is significant only at ZD 55°

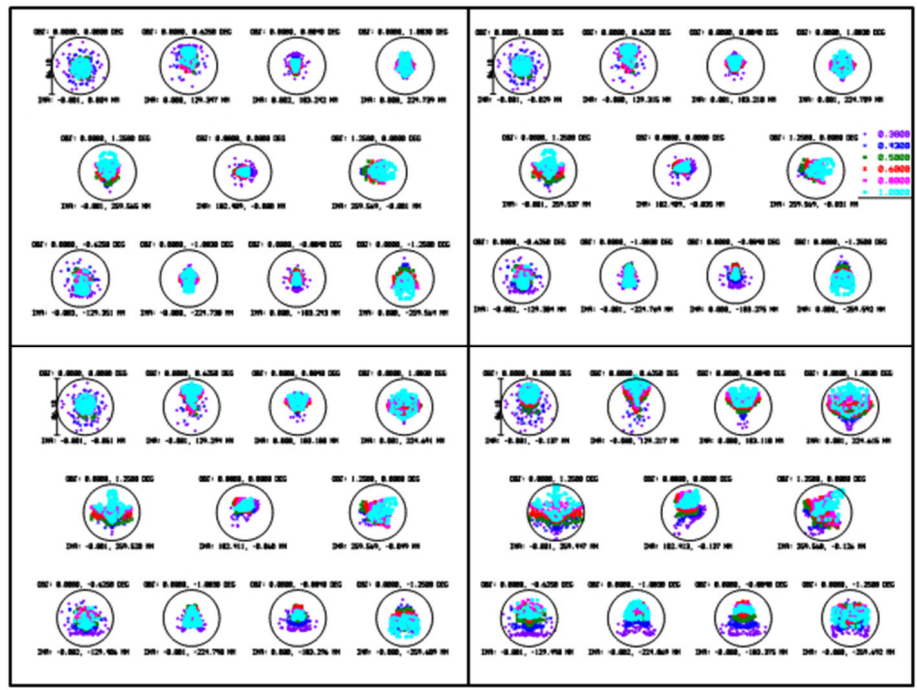

Figure 8. Spot diagrams at four zenith angles with ADC mechanism DS. Top left: 0, top right: 20°, bottom left 38°, and bottom right 55°. The circle diameters are equivalent to 1.5 arcsec.

## 2.4 Effect of temperature

The atmospheric dispersion changes with air temperature, pressure, and humidity but, for conditions suitable for observing, only the variation with temperature is significant. For a change of ±10°C, the optimum ADC offset at ZD 55° changes by ±0.34 mm (in the sense of requiring larger offset for lower temperature).

The offset at a given zenith angle will vary with temperature primarily through change in the elastic modulus of the spring material. From the experimental results reported for a stainless steel tuning fork, the temperature coefficient of the Youngs Modulus E was $-2.7 \times 10^{-4}$/°C. Assuming the shear modulus has the same temperature coefficient, the extension of a spring for a given load will increase by 0.27% for a 10°C rise in temperature. This is in the opposite sense to that needed to compensate the seasonal change but it increases the seasonal offset effect only from ±0.34mm to ±0.36mm

These errors are very small compared with the ADC displacement, equivalent to the ADC being set for a ZD incorrect by only ± 2° at ZD 55°.

## 2.5 Check on distortion due to ADC action

In some designs, the ADC introduces sufficient asymmetry in distortion across the field that the change in this distortion during an exposure of 30 minutes duration can be comparable with or worse than the typical change in differential atmospheric refraction across the field in the same time. This compromises the application of the ADC and can lead to the strategy of setting the ADC at the optimum for the mid-point of the exposure and not changing it during the exposure. Such a strategy would, of course, not be possible with a purely passive drive of the ADC. So it is important to confirm that the current design does not have this limitation.

To quantify the effect, insofar as it would cause trailing of an image in the field center if guide stars were used at the field periphery, the ray tracing includes tilt of the telescope as a whole to place the chief ray from the on axis star (that with X field and Y field both = 0) at the center of the field. Then the Y positions in the focal surface of the images for the stars at the positive and negative extremes in Y field are checked for the various ZD settings. Table 1 shows the results for the current design.

Table 1. $Y_{centroid}$ (mm) in focal surface and changes in Y with ZD setting.

| Y(Xfield,Yfield): | Y(0, 0) | Y(0, 1.25°) | Y(0, -1.25°) | Mean $Y_{1.25, -1.25}$ | Mean – Y(0,0) µm |
|---|---|---|---|---|---|
| ZD 0 | 0.000 | 259.565 | -259.565 | 0.000 mm | 0 |
| ZD 20° | 0.002 | 259.568 | -259.562 | 0.003 | 1 |
| ZD 38° | 0.003 | 259.572 | -259.559 | 0.0065 | 3.5 |
| ZD 55° | 0.006 | 259.578 | -259.555 | 0.0115 | 5.5 |

The mean of the Y positions of the images of the two stars at the Yfield extremes changes in relation to the image of the on-axis star by only 5.5 µm due to the ADC induced distortion from ZD 0 to ZD 55°. At the latitude of VISTA, during any exposure of 30 minutes, the ZD cannot change more than 7° so the change in differential distortion cannot be more than ~1 µm, which is inconsequential compared with typical changes in differential atmospheric refraction for the same change in ZD.

So there is no observational disadvantage in having the ADC setting completely automatic and passive.

# 3. MECHANICS OF ADC OFF-SETTING

## 3.1 Simple case with deflection proportional to sin(ZD)

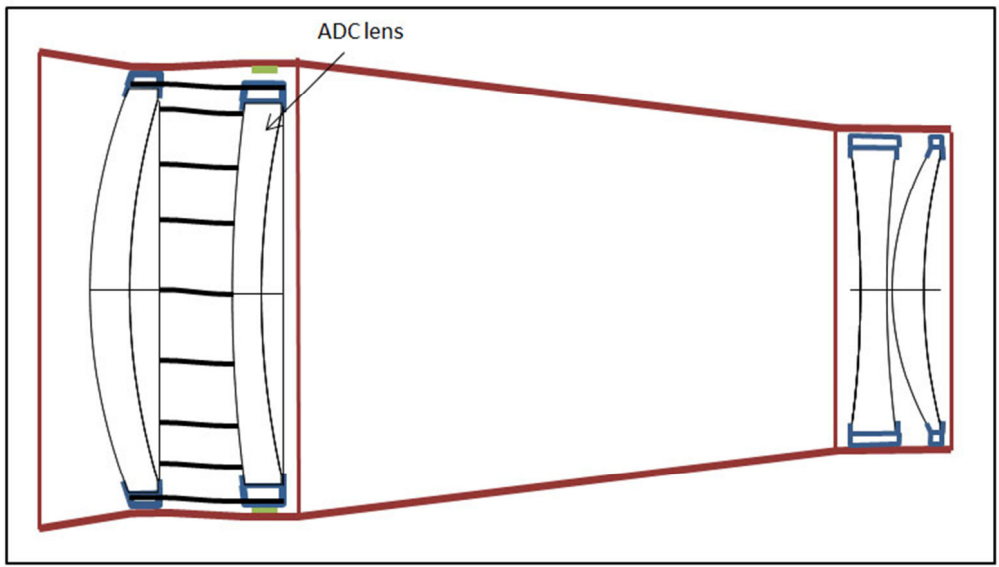

Figure 9. Sketch of the way in which the ADC element could be supported with multiple flex rods so as to be offset through a distance proportional to sin(ZD) (restraint arrangement **SS**).

Figures 9, and 10 show roughly how the simpler off-setting, proportional to sin (ZD), can be implemented. 16 flexure rods (of stainless spring steel) 5.9 mm in diameter and with effective length 300 mm give the appropriate lateral compliance and are stressed in bending to a maximum of about 1/4 of yield point. At the zenith they will have an insignificant tensile stress and they have a huge factor of safety against buckling as columns if the corrector assembly was inverted (to point downwards) during handling.

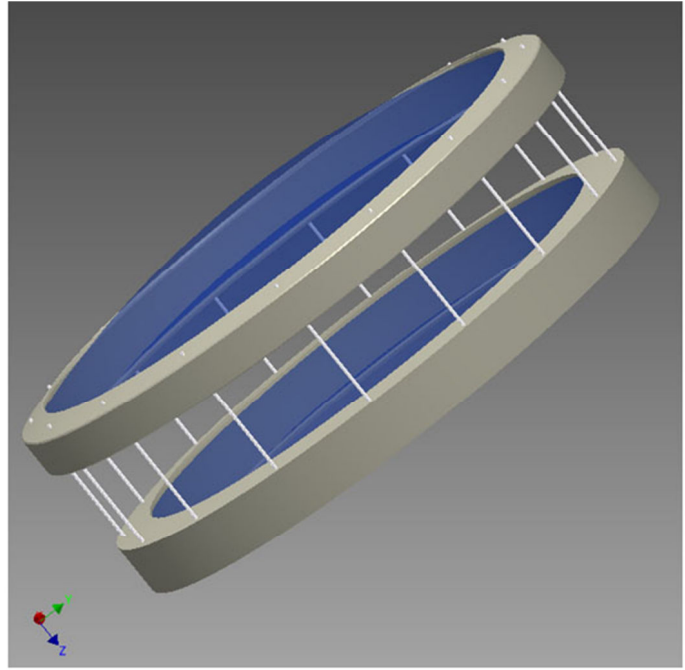

Figure 10 CAD model of elements 1 and 2 with 16 flex rods connecting the cells so that the ADC (element 2) will translate in the negative Y direction through a distance proportional to sin(ZD).

## 3.2 Case with deflection close to tan(ZD) proportionality

The essence of the proposed mechanism is illustrated in figure 11. For two opposite pre-tensioned springs, each with a spring constant K N/m, it results in lateral deflection proportional to sin (ZD)/2K for part of the ZD range, from 0 to say 30°, and proportional to sin (ZD)/K for greater ZD. This allows a good fit to the tan function.

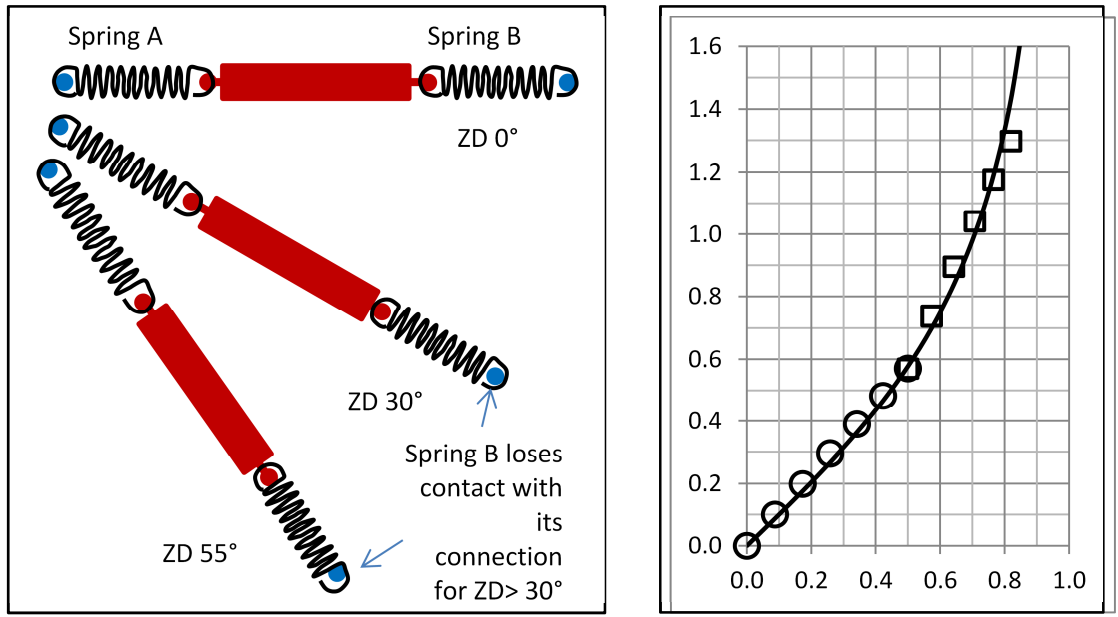

Figure 11 Left: Illustration of the principle of two stage lateral restraint of the ADC. The red rectangle represents the ADC and its cell; the blue circles represent anchors to the main lens cell structure. At the zenith, springs A and B, each having a spring constant **K** (N/m), are under equal tension. At ZD 30°, the tension in spring B goes to zero and for any larger ZD, only spring A is changing in length. So, up to ZD 30°, the effective restraining force is proportional to **2Kd** where d is the lateral displacement; beyond 30° the spring rate is halved, to **Kd**.
Right: Result of two stage restraint for two springs 180° apart around the lens cell, as shown on the left. The curve is tan θ v sin θ. The circles and squares are points on two straight lines with slopes in the ratio 1:2. The points are at intervals of 5° in theta from 0 to 55°.

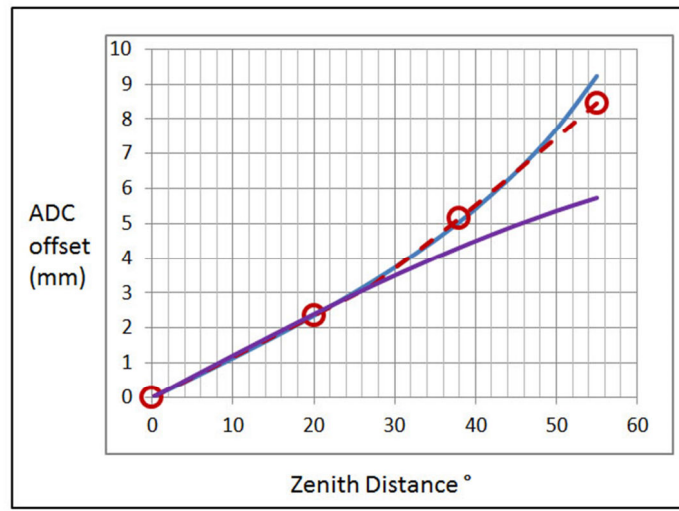

Figure 12 Effect of allowance for summation of spring forces with 16 springs around the cell periphery. The lower curve is proportional to sin(ZD) the upper curve, to tan(ZD) and the dashed curve is the function obtained with the 16 springs. The circles are at the values adopted for the ray-tracing at 0, 20°, 38°, and 55°.

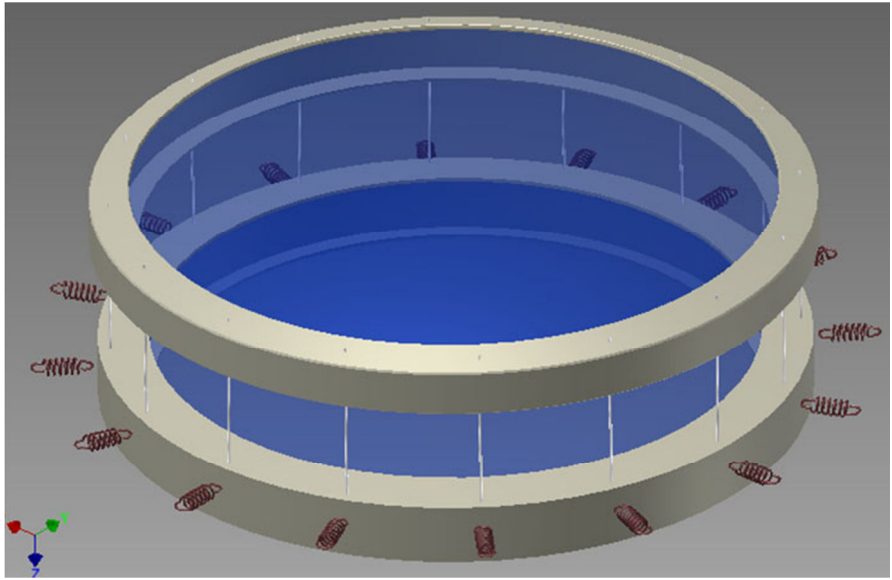

Figure 13 CAD model with tension springs in correct proportion, suggesting how they would be used together with thin flex rods to constrain the ADC cell to deflect laterally in two stages.

Although VISTA has an alt-azimuth mounting, the provision for supporting the corrector lens is to attach it to the Cassegrain instrument rotator; so the corrector position angle with respect to gravity will go through a full circle. Unfortunately, this complicates the challenge of modifying the ADC restraint to a near tan(ZD) function. To operate equally well at any position angle with respect to gravity requires that the opposing pair of springs be repeated at several position angles around the periphery of the ADC cell. When a calculation is made of the combined result with 16 springs equally spaced around the cell, the 2:1 ratio in deflection rates is reduced to 1.57:1. This still provides a much better match to the tan function than a single sine function, as indicated in figure 12.

Figure 13 indicates the proportions of appropriate helical tension springs in relation to the ADC cell. Thinner flex rods (3 mm diameter, 300 mm long) than those for the simpler mounting would couple the ADC cell to the element 1 cell, since they must not contribute significantly to the lateral stiffness of the arrangement. With these dimensions, the safety factor against buckling if the corrector was inverted would be less than with the 5.9 mm diameter rods but still adequate – about a factor 8.

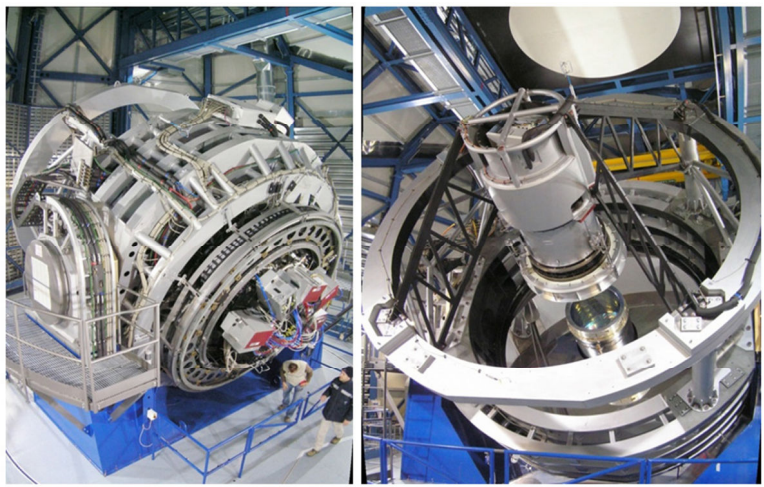

Figure 14 Views of VISTA (in its IR mode) to emphasise the desirability of keeping the 4MOST corrector ADC implementation simple and maintenance-free.

# 4. SPECIAL ADVANTAGES OF THIS CORRECTOR ON VISTA

Figure 14 shows two views of VISTA to emphasize the advantage of minimizing the complexity of any mechanisms associated with the corrector. For the 2012 design, the two ADC doublets, their bearings, gearing, motors, and encoders would be buried between the complex Cassegrain-mounted multi-fiber positioner and the primary mirror, with very poor access for any maintenance. For the present design, the single element ADC, with none of these components, is near the front end of the corrector, clear of M1.

## 4.1 Summary of optical and other advantages of this design

The new design offers the following advantages over the October 2012 design:

1. Significantly smaller images, giving higher efficiency in feeding optical fibres.
2. Improved transmission, especially toward the UV end of the design range.
3. An ADC that can be mounted more simply than a pair of rotating ADC components and is driven entirely by gravity with no need for motors, encoders, cabling or software control.
4. No need for space behind M1 to mount ADC drive gearing, motors and encoders for two rotating doublets.
5. 4 elements rather than 6, reducing the optical manufacturing cost.
6. About 23% less finished mass of glass, reducing the material cost and simplifying mechanical support of the elements.
7. No need for cementing or oiling together large lens elements.
8. Nearly 4% faster focal ratio: f/3.20 rather than f/3.32. This allows slightly less tilt of the positioner spines for a given angular patrol radius, improving light collection efficiency.

The only disadvantage is the increase in the number of aspheric surfaces from two to four.